\documentclass[manuscript]{aastex}
\usepackage{gensymb}

\shorttitle{Ice Regulated Chemistry}
\shortauthors{}

\begin{document}

\title{H$_2$CO and N$_2$H$^+$ in Protoplanetary Disks: 
Evidence for a CO-ice Regulated Chemistry}

\author{Chunhua Qi}

\affil{Harvard-Smithsonian Center for Astrophysics, 60 Garden Street, 
Cambridge, MA 02138, USA} 

\and

\author{Karin I. \"Oberg}

\affil{University of Virginia, Departments of Chemistry and Astronomy, Charlottesville, VA 22904, USA}

\and

\author{David J. Wilner}

\affil{Harvard-Smithsonian Center for Astrophysics, 60 Garden Street, 
Cambridge, MA 02138, USA}

\begin{abstract}

We present Submillimeter Array observations of H$_2$CO and N$_2$H$^+$
emission in the disks around the T Tauri star TW~Hya and the Herbig Ae star
HD~163296 at 2$''$--6$''$ resolution and discuss the distribution of these species
with respect to CO freeze-out.
The H$_2$CO and N$_2$H$^+$ emission toward HD~163296 does not peak at the
continuum emission center that marks the stellar position but is instead
significantly offset. Using a previously developed model for the physical
structure of this disk, we show that the H$_2$CO observations are reproduced
if H$_2$CO is present predominantly in the cold outer disk regions. A model
where H$_2$CO is present only beyond the CO snow line
(estimated at a radius of 160~AU) matches the observations well.
We also show that the average H$_2$CO excitation temperature,
calculated from two transitions of
H$_2$CO observed in these two disks and a larger sample of
disks around T Tauri stars in the DISCS (the Disk Imaging Survey of
Chemistry with SMA) program,
is consistent with the CO freeze-out temperature of $\sim$20~K.
In addition, we show that N$_2$H$^+$ and H$_2$CO line fluxes in disks are
strongly correlated, indicative of co-formation of these species across the
sample. Taken together, these results imply that H$_2$CO and N$_2$H$^+$ are
generally present in disks only at low temperatures where CO depletes onto
grains, consistent with fast destruction of N$_2$H$^+$ by gas-phase CO, and
{\it in situ} formation of H$_2$CO through hydrogenation of CO ice.
In this scenario H$_2$CO, CH$_3$OH and N$_2$H$^+$ emission in disks should
appear as rings with the inner edge at the CO midplane snow line.
This prediction can be tested directly using observations from ALMA with higher
resolution and better sensitivity.

\end{abstract}


\section{Introduction}
Planetary systems are assembled from dust and gas in the disks
surrounding pre-main sequence stars. The nature of the formed planets
are intimately linked to the structure, composition and evolution of
the parent circumstellar disk. Molecular emission lines serve as probes 
of disk characteristics, such as density, temperature and ionization 
fraction, that are not accessible by other observations. For example,
N$_2$H$^+$ along with the deuterated ions like H$_2$D$^+$ and DCO$^+$ 
are believed to trace the ionization fraction near the midplane of the 
disks \citep{Oberg11d}. 
Molecular distributions in disks are also important to characterize 
because of their connection with the composition of forming planetesimals. 
This especially true of organic molecules, of which H$_2$CO is an 
important representative. While most molecules are expected to be 
reprocessed in larger planetary bodies, the disk-chemical composition 
may survive quite intact in icy planetesimals, including comets \citep{Mumma11}. Such planetesimals 
may have seeded the Earth with water and organics, connecting disk 
chemistry with the origins of life. Predicting the organic composition 
of these planetesimals depend on our understanding of the distribution of 
the organic composition of grains in the disks.

The overall disk chemical structure is set by a combination of 
photochemistry at the disk surface and sequential freeze-out of molecules 
in the disk interior \citep[e.g.][]{Aikawa02}. The distributions of N$_2$H$^+$ 
and H$_2$CO and their relationship to CO present important test cases
of the chemical models. The CO molecule is one of the last to freeze out 
and predicted to deplete quickly from the gas-phase at T$<20-25$~K 
for typical disk mid-plane densities. Abundant N$_2$H$^+$ is expected 
only where CO is depleted because N$_2$H$^+$ forms from protonation of 
N$_2$, which remains in the gas-phase at temperatures a few degrees 
lower than CO \citep{Oberg05}, and is destroyed mainly by reactions 
with CO \citep{Bergin02}. In dense cores in star
forming regions, the abundance of CO is observed to 
show a strong anti-correlation with N$_2$H$^+$
\citep[e.g.][]{Caselli99, Bergin02, Jorgensen04}. 
In disk models, this effect is manifested
as a jump in the N$_2$H$^+$ column density at the CO ``snow line'',
one order of magnitude in a recent calculation \citep{Walsh12}.
The snow line is here defined as
the disk radius where the midplane dust temperature is cold enough for
volatiles to condense into ice grains. 

The H$_2$CO chemistry is more complicated than N$_2$H$^+$ 
because H$_2$CO can form through several different pathways, both in the gas-phase and on grain-surfaces. Grain-surface formation of H$_2$CO should depend directly on CO freeze-out; constrained by theory \citep{Tielens82,Cuppen09} and experiments
\citep[e.g.][]{Watanabe03,Fuchs09}, H$_2$CO
(and CH$_3$OH) form readily from CO ice hydrogenation. If this is 
the dominant formation pathway of H$_2$CO in disks, and if ices are
partially desorbed non-thermally
\citep[e.g.][]{Garrod07,Oberg09b,Oberg09c}, then H$_2$CO gas should 
coincide with N$_2$H$^+$ exterior to  the CO snow line. 

To date, emission from millimeter wavelength H$_2$CO and N$_2$H$^+$ lines 
has been detected toward 8 and 6 protoplanetary disks, respectively
\citep{Dutrey97,Aikawa03,Qi03,Thi04,Dutrey07,Henning08,Oberg10c,Oberg11a}. 
Most detections are toward T Tauri stars with massive disks, and
the detection fraction toward more luminous Herbig Ae stars is low
\citep{Oberg11a}. Based on these observations, H$_2$CO and N$_2$H$^+$
are mainly abundant in disks with large reservoirs of cold dust and gas,
where CO freeze-out is expected to occur. A direct connection between CO freeze-out and N$_2$H$^+$ and H$_2$CO in disks has yet to be observationally established, however.  
   
In this paper we present Submillimeter Array (SMA) observations of
H$_2$CO and N$_2$H$^+$ toward the disks around HD~163296 and TW~Hya, 
and we use these new observations along with H$_2$CO and N$_2$H$^+$
observations from DISCS (Disk Imaging Survey of Chemistry with SMA)
to constrain the H$_2$CO and N$_2$H$^+$ distributions.
The new data and their calibration are described in \S\ref{sec:obs}. 
In \S\ref{sec:res}, we present the H$_2$CO and N$_2$H$^+$ images and 
spectra toward HD~163296 and TW~Hya, models of the H$_2$CO distribution 
toward HD~163296, H$_2$CO excitation temperature calculations, and 
examine the relationship between H$_2$CO and N$_2$H$^+$ emission 
across the sample of disks. 
In \S\ref{sec:disc}, we discuss the implications of these results,
summarize the mounting evidence for CO-ice 
regulated chemistry and make predictions for future observations of 
H$_2$CO and N$_2$H$^+$ emission from disks with better sensitivity and 
resolution.
 
\section{Observations\label{sec:obs}}

The observations of HD~163296 
(${\rm R.A.=17^h56^m21.279^s}$, 
${\rm decl. = -21^\circ57\arcmin22\farcs38}$; J2000.0) 
were made between 2008 and 2012, 
and of TW Hydrae (${\rm R.A. = 11^h01^m51.875^s}$,
${\rm decl. = -34^\circ42\arcmin17\farcs155}$; J2000.0) 
between 2008 and 2012, using the eight-antenna Submillimeter Array
(SMA) located atop Mauna Kea, Hawaii. 
Table \ref{tab:obs} provides a summary of the observational parameters 
and results. 
For the 2007 and 2008 observations, the SMA receivers operated 
in a double-sideband mode with an intermediate frequency (IF) band of 
4--6 GHz from the local oscillator frequency, sent over fiber optic 
transmission lines to 24 overlapping ``chunks'' of the digital correlator. 
The 2012 observations were made after an upgrade that enabled a second 
IF band of 6--8 GHz, effectively doubling the bandwidth.

The SMA observations of HD~163296 were carried out in the compact-north 
(COM-N), compact (COM) and subcompact (SUB) array configurations. 
The 2007 observations included the DCO$^+$ 3--2 line at 216.1126 GHz
and the H$_2$CO $3_{1,2}-2_{1,1}$ line at 225.698 GHz.
The 2012 observations included the N$_2$H$^+$ 3--2 at 279.512 GHz
and the H$_2$CO $4_{1,4}-3_{1,3}$ line at 281.527 GHz.
The observing loops used J1733--130 as the main gain calibrator
and observed J1744--312 every other cycle to check the phase calibration.
Flux calibration was done using observations of Titan and Uranus. 
The derived fluxes of J1733-130 were 
1.19 Jy (2007 Mar 20), 1.25 Jy (2012 Jun 10), 1.40 Jy (2012 Aug 12 and 14). 
The bandpass response was calibrated using observations of 3C279, Uranus 
and J1924--292. 

The SMA observations of TW~Hya were carried out in the compact (COM) and 
subcompact (SUB) array configurations. 
The 2008 observations included the H$_2$CO $5_{1,5}-4_{1,4}$ line at 351.769 GHz. 
The 2012 Jan 13 SUB observation included the H$_2$CO $4_{1,4}-3_{1,3}$ and N$_2$H$^+$ 3--2 lines, 
like HD~163296, but unfortunately the chunk containing N$_2$H$^+$ 
was corrupted and unusable. 
The 2012 Jun 04 COM observation using a similar setting successfully included the N$_2$H$^+$ line. The observing loops used J1037--295 as the
gain calibrator. Flux calibration was done using observations of Titan 
and Callisto.  The derived fluxes of J1037--295 were 0.73 Jy (2008 Feb 23), 
0.73 Jy (2012 Jan 13) and 0.82 Jy (2012 Jun 4). The bandpass response was 
calibrated using observations of 3C279 and 3C273. 

Routine calibration tasks were performed using the MIR software
package \footnote{http://www.cfa.harvard.edu/$\sim$cqi/mircook.html}, 
and imaging and deconvolution were accomplished in the MIRIAD software package.

\section{Results}\label{sec:res}

In this section we present detections of H$_2$CO and N$_2$H$^+$ 
emission lines toward HD~163296 and TW~Hya, display their respective 
distributions (\S\ref{det}), and compare the higher quality observations 
of H$_2$CO in HD~163296 with models (\S\ref{model}).
We then combine the new data with previously reported H$_2$CO and 
N$_2$H$^+$ detections in disks to examine trends with respect to 
H$_2$CO excitation temperature (\S\ref{temp}) and with each other 
(\S\ref{corr}). 

\subsection{H$_2$CO and N$_2$H$^+$ towards HD~163296 and TW~Hya\label{det}}

\begin{figure}[htp]
\epsscale{1.0}
\plotone{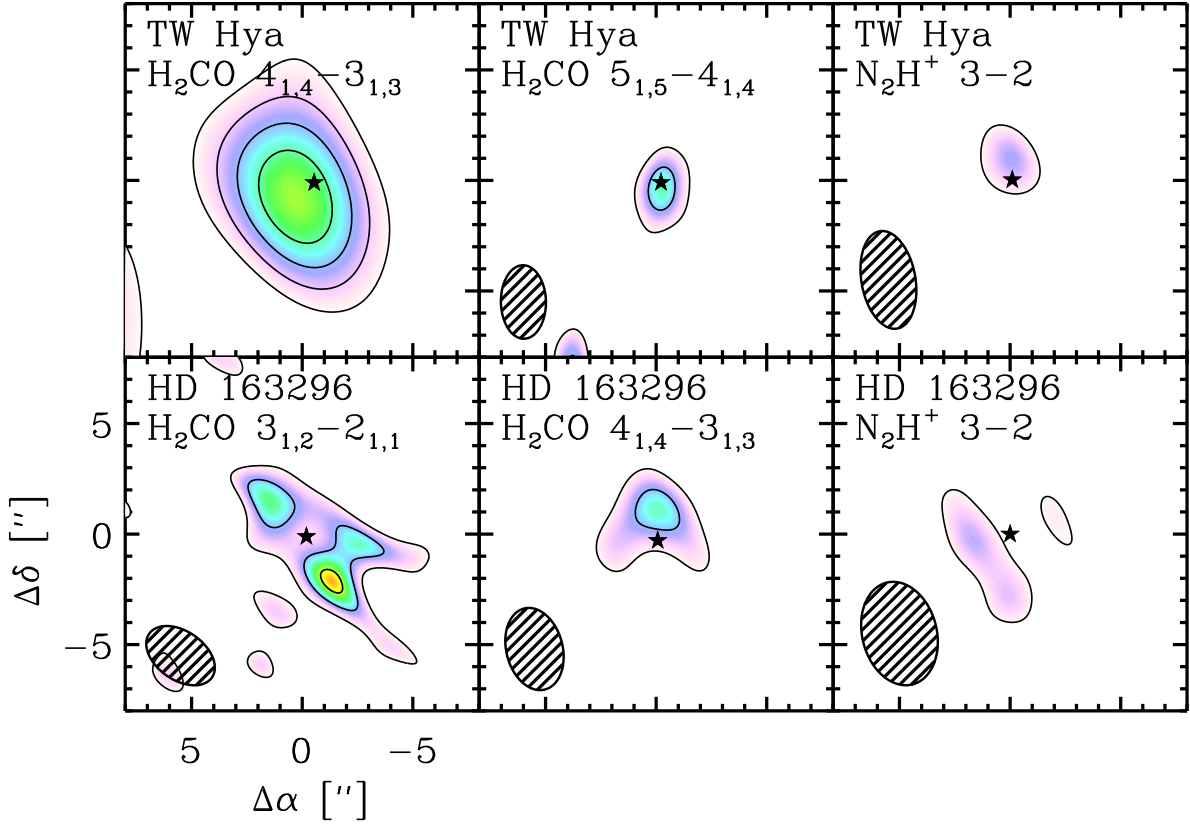}
\caption{H$_2$CO and N$_2$H$^+$ observations toward TW~Hya and HD~163296 with the SMA. The emission toward TW~Hya is centrally peaked, while there is a clear offset in the H$_2$CO emission toward HD~163296. The first two contour levels are 3 and 5$\sigma$, with 1$\sigma$ measured to be 0.10 (H$_2$CO 4--3),  0.13 (H$_2$CO 5--4), and 0.36 (N$_2$H$^+$) Jy km s$^{-1}$ per beam toward TW~Hya, and 0.06 (H$_2$CO 3--2),  0.17 (H$_2$CO 4--3), and 0.13 (N$_2$H$^+$) Jy km s$^{-1}$ per beam toward HD~163296. \label{fig:obs_mom}}
\end{figure}

Figure \ref{fig:obs_mom} shows images of the spectrally integrated 
emission toward TW~Hya and HD~163296 at the rest frequencies of two H$_2$CO 
lines and the N$_2$H$^+$ $J=3-2$ line. 
H$_2$CO $4_{1,4}-3_{1,3}$ and $5_{1,5}-4_{1,4}$ and N$_2$H$^+$ are 
detected toward TW~Hya, and H$_2$CO $3_{1,2}-2_{1,1}$ and 
H$_2$CO $4_{1,4}-3_{1,3}$  and N$_2$H$^+$ are detected toward HD~163296.
The emission toward TW~Hya appears to be centrally peaked 
at the size scale of the beams (FWHM $>2\arcsec$).

Toward HD~163296, however, neither the H$_2$CO lines nor the N$_2$H$^+$ line 
emission peaks at the location of the continuum peak
that marks the stellar position, but instead show significant offsets. 
This is most readily apparent for the  H$_2$CO $3_{1,2}-2_{1,1}$ line that 
was observed with a slightly smaller and more advantageously rotated beam, 
where the emission appears ring-like. This is the second reported observation 
of a ring-like H$_2$CO distribution after DM~Tau \citep{Henning08}. Interpreting the H$_2$CO emission toward DM~Tau is however complicated by a large central cavity in dust emission \citep{Andrews11}. 

\begin{figure}[htp]
\epsscale{1.0}
\plotone{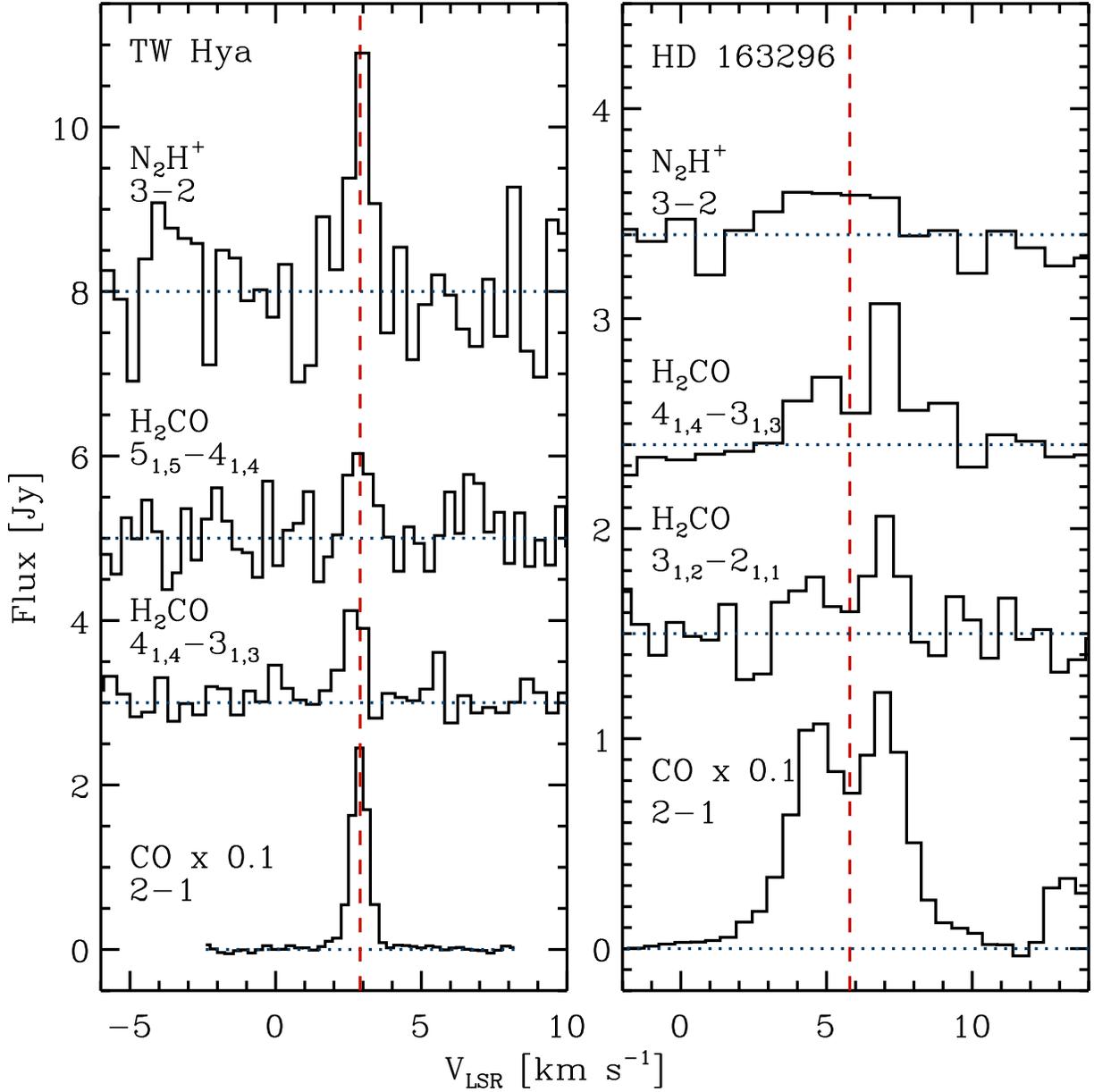}
\caption{Spatially integrated spectra of H$_2$CO and N$_2$H$^+$ toward TW~Hya and HD~163296. The red dashed lines mark V$_{\rm LSR}$ toward each source, based on CO observations \citep{Qi06,Qi11}.  The double-peak structure typical for rotating disks is not resolved toward TW~Hya with the applied spectral resolution because of its face-on orientation. \label{fig:obs_spec}}
\end{figure}

Figure \ref{fig:obs_spec} shows the spatially integrated spectra. 
The line shapes and central velocities agree with what has been previously 
observed for other molecular lines toward these disks. 
The line fluxes are listed in Table \ref{tab:obs}, and the values are 
comparable to detections of these lines toward other large protoplanetary 
disks \citep{Oberg10c,Oberg11a}. 

\subsection{HD~163296 H$_2$CO Model Results \label{model}}

Figure \ref{fig:obs_mom} demonstrates that H$_2$CO emission toward
HD~163296 is spatially resolved and thus contains information on the
radial distribution of H$_2$CO in the disk. The relative excitation of the two 
H$_2$CO transitions should probe primarily the vertical distribution and thus provide complementary constraints. 
We explore the H$_2$CO distribution based on a previously developed
accretion disk model with a well-defined temperature and density 
structure, constrained by the HD~163296 broadband spectral energy 
distribution, spatially resolved millimeter dust continuum, and 
multiple CO and CO isotopologue line observations \citep{Qi11}. 
We adopt the same methods as \citet{Qi08} for constraining the 
H$_2$CO distribution, here fitting models that assume a radial 
power-law (\S\ref{power}) and a simple ring with inner boundary at the 
CO ``snow line'' (\S\ref{ring}).  

\subsubsection{Power-law Model \label{power}}

For a first-order analysis of the distribution of H$_2$CO, 
we model the radial variation in the column density as a power law 
N$_{100}\times(r/100)^p$ between an inner radius R$_{in}$ and outer radius
R$_{out}$, where N$_{100}$ is the column density at
100 AU in cm$^{-2}$, $r$ is the distance from the star in AU, and $p$ 
is the power-law index. 

For the vertical distribution, we assume that H$_2$CO is present 
with a constant abundance in a layer with boundaries toward the 
midplane and toward the surface of the disk (similar to \citet{Qi08}).
This assumption is motivated by 
chemical models \citep[e.g.][]{Aikawa06} that predict a three-layered 
structure where most molecules are photodissociated in the surface layer, 
frozen out in the midplane, and have an abundance that peaks at 
intermediate disk heights. The surface ($\sigma_s$) and midplane 
($\sigma_m$) boundaries are presented in terms of 
$\Sigma_{21}=\Sigma_H/(1.59\times10^{21} cm^{-2})$, where $\Sigma_H$
is the hydrogen column density measured from the disk surface. 
This simple model approximates the vertical location where H$_2$CO is 
most abundant.  The excitation of multiple transitions can constrain 
both $\sigma_s$ and $\sigma_m$, but in this case of very modest 
signal-to-noise, we fix $\sigma_s$ to 0.79, the surface boundary found
for CO by \citet{Qi11}, and we fit $\sigma_m$ for the midplane
boundary and the power-law parameters (N$_{100}$, $p$, R$_{in}$ and R$_{out}$). 

Using the structure model, we 
compute a grid of synthetic H$_2$CO visibility datasets over a
range of R$_{out}$, R$_{in}$, $p$, $\sigma_m$ and N$_{100}$ values and compare
with the observations. The best-fit model is obtained by minimizing
$\chi^2$, the weighted  
difference between the real and imaginary part of the complex visibility 
measured in the ($u,v$)-plane sampled by the SMA observations of 
both H$_2$CO transitions. 
We use the two-dimensional Monte Carlo model RATRAN \citep{Hogerheijde00} 
to calculate the radiative transfer and molecular excitation. 
The collisional rates are taken from the Leiden Atomic and Molecular 
Database \citep{Schoier05}.  

\begin{figure*}[htp]
\epsscale{1.0}
\plotone{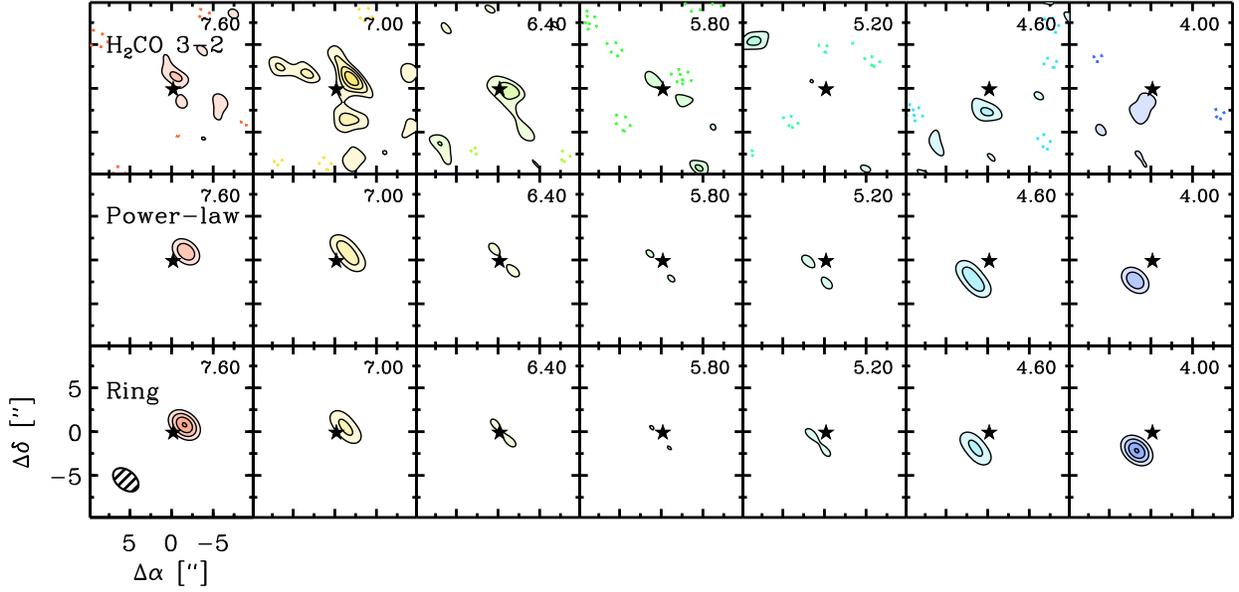}
\caption{Observed and simulated channel maps for H$_2$CO
  $3_{1,2}-2_{1,1}$ toward HD~163296, using the best-fit power-law (p=2) and ring
  model. Both models fit the data within the uncertainties. 
The first contour is 2$\sigma$ and each following contour step is 1$\sigma$. The channel velocity in km s$^{-1}$ is in the upper right corner of each panel and the synthesized beam is displayed in the lower left panel. \label{fig:32_chmap}}
\end{figure*}

\begin{figure*}[htp]
\epsscale{1.0}
\plotone{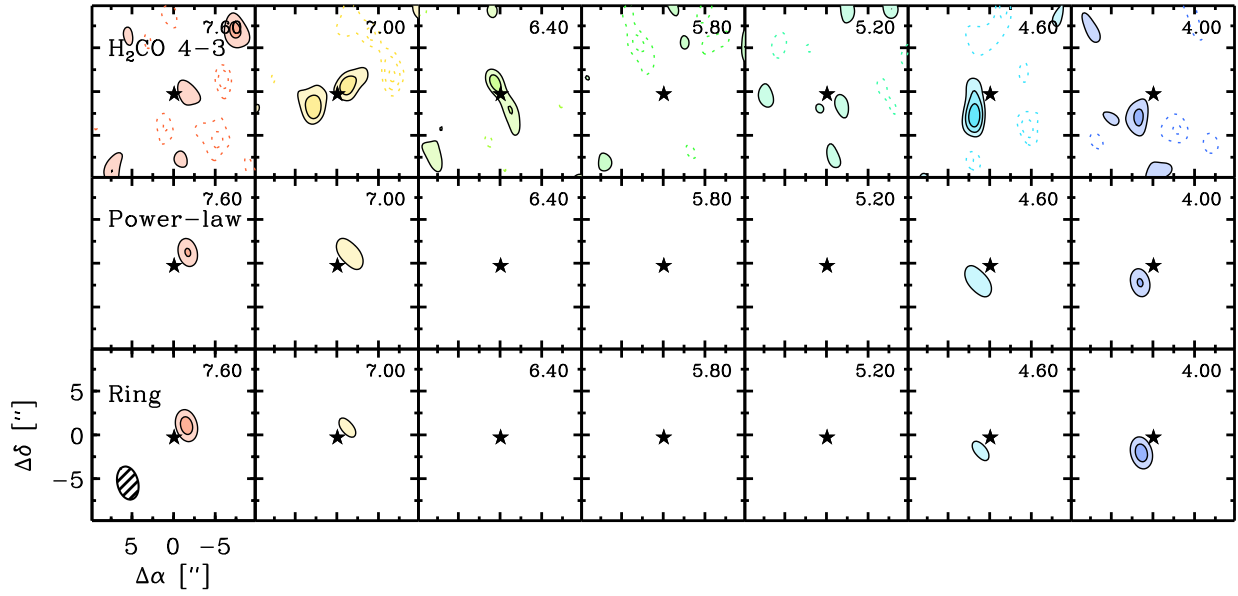}
\caption{As Fig. \ref{fig:32_chmap} but for the H$_2$CO $4_{1,4}-3_{1,3}$ line.  \label{fig:43_chmap}}
\end{figure*}

Table~\ref{tab:best-fit} lists the best-fit parameters of the model.
The power-law index of 2 implies an H$_2$CO column density that strongly
increases with radius. Figure \ref{fig:model} shows the best-fit radial 
distribution of the H$_2$CO column density. 
Figures \ref{fig:32_chmap} and \ref{fig:43_chmap} present comparisons 
between the observed channel maps and the best-fit model. The model 
reproduces the main features of the observations remarkably well, 
in particular the flux ratio between the inner and outer channels, 
and the lack of emission at the location of the continuum peak. 

\begin{figure}
\epsscale{0.6}
\plotone{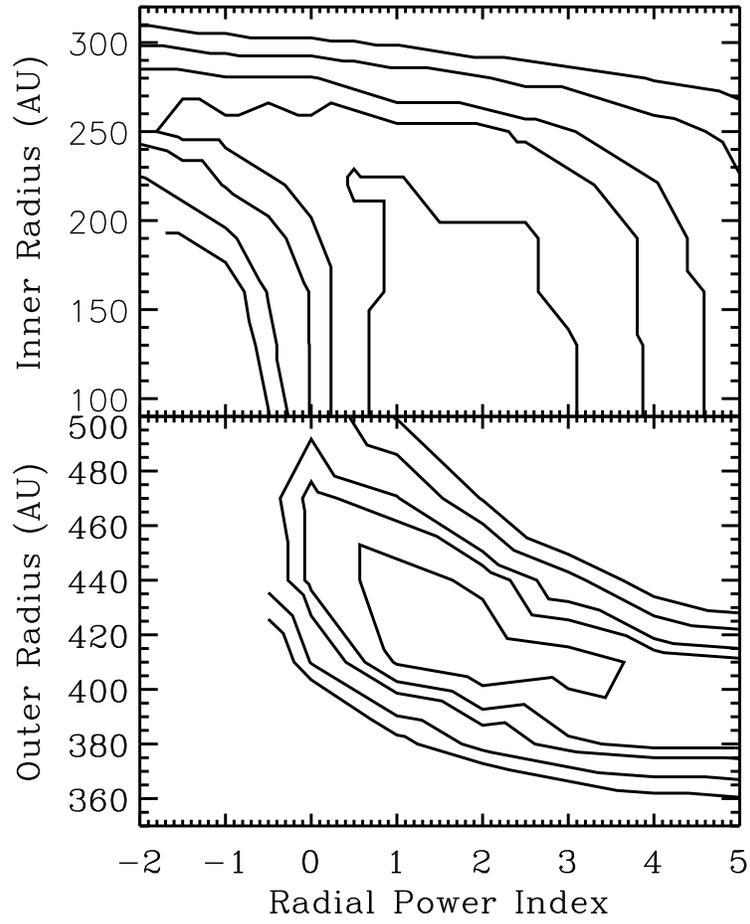}
\caption{Iso-$\chi^2$ surfaces of $R_{\rm out}$ and $R_{\rm in}$
  versus $p$. Contours correspond to the 1--5 $\sigma$ errors. 
\label{fig:chi2}}
\end{figure}

\begin{figure}
\epsscale{1.0}
\plotone{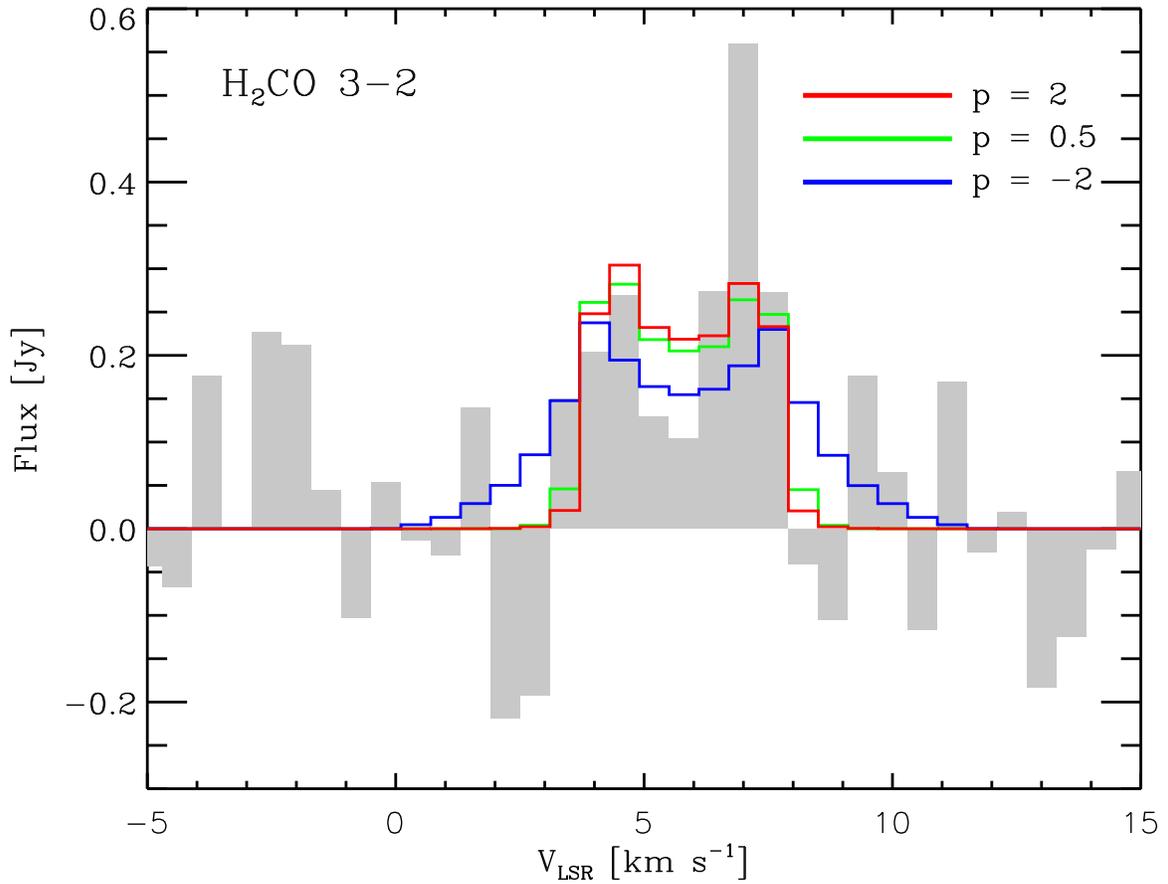}
\caption{Simulated spectra for H$_2$CO
  $3_{1,2}-2_{1,1}$ for models with radial column densities power-law
  indices $p=2$(best-fit, red line), $0.5$(within 1$\sigma$ noise level, green line) and $-2$(blue line), overlaid
  with the HD 163296 spectra in gray shade.  
\label{fig:mod_spec}}
\end{figure}

Figure~\ref{fig:chi2} shows the $\chi^2$ surfaces for the R$_{in}$ and
R$_{out}$ versus the power law index $p$, which enables us to quantify the
uncertainties associated with the inner and outer region sizes and the
power-law index.
We find that $p$ is constrained between 0.5--3.0 (within 1$\sigma$) while
R$_{in}$ is constrained to be $<$200 AU. The $\chi^2$ value does not change
significantly for inner radii $<$90 AU, as expected for the 2$\arcsec$
beam size of the observations. The outer radius is better constrained, since
the emission is very sensitive to the value of R$_{out}$ with a positive
power-law index. Figure~\ref{fig:mod_spec} shows the H$_2$CO 3--2 line
spectrum compared with the spectra derived from models with different
radial column densities power-law indices. The spectra suggest a lack of
high velocity line wings associated with emission originating in the inner
regions of the disk, consistent with the results of the $\chi^2$ analysis.
  
\subsubsection{Ring Model \label{ring}}

The positive power-law index found for the H$_2$CO radial distribution 
implies that H$_2$CO is present mainly in the outer disk. 
This is expected if H$_2$CO forms {\it in situ} from CO ice hydrogenation 
and is therefore present mainly beyond the CO snow-line, previously
determined to be at 160~AU by \cite{Qi11}.
Guided by this astrochemical {\it ansatz}, we have tried a second ``ring'' 
model where the H$_2$CO gas is only present where CO has frozen out. 
The vertical surface boundary is then defined by the CO freeze-out 
temperature of 19~K \citep{Qi11}, while the midplane boundary can be
constrained by the excitation of multiple H$_2$CO transitions, as in
the power-law model. Within this layer, the abundance of 
H$_2$CO is assumed to follow that of H nuclei with a constant fractional 
abundance, which is also a parameter fit to the data. 

\begin{figure}[htp]
\epsscale{1.0}
\plotone{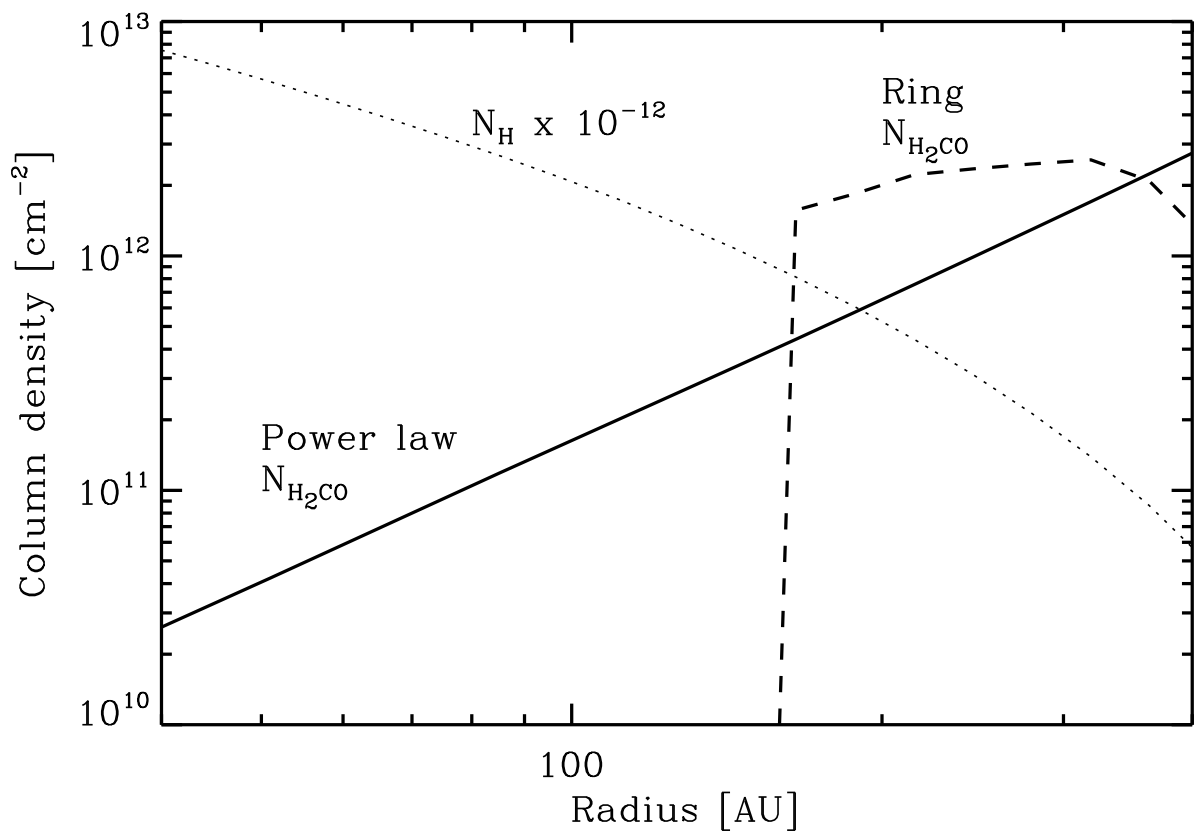}
\caption{The hydrogen nuclei column density toward HD~163296
  \citep{Qi11} is plotted together with the H$_2$CO column density
  from the best-fit power law and `ring' models. In the ring model,
  the H$_2$CO abundance is defined to be zero when the temperature is
  less than 19 K, the CO freeze-out temperature, which leads to the
  sharp drop in the H$_2$CO column density interior to the CO snow
  line at 160 AU. \label{fig:model}} 
\end{figure}

The best-fit abundance is $5.5\times10^{-11}$ and the midplane
boundary $\sigma_m$ is consistent with what we find in the power-law
model (Table~\ref{tab:best-fit}). 
Figure \ref{fig:model} shows that the vertical surface boundary at 19~K 
effectively results in a ring-like radial structure, where the inner edge 
of the ring is at CO snow line. Figure \ref{fig:model} also shows that the 
best-fit power-law and ring models result in similar H$_2$CO column densities  
beyond the CO snow line. The profile of the ring model is considerably 
flatter than the power-law model and even drops outside of 300~AU
exponentially to the edge of CO emission. 
Figures \ref{fig:32_chmap}--\ref{fig:43_chmap} show that the power-law 
model and the ring model channel maps display some subtle differences. But 
both provide good fits to the data within the noise of the SMA observations.
The same model also provides a good match to the N$_2$H$^+$ flux, but the 
combination of low signal-to-noise and observations of just one transition 
preclude any independent modeling of the N$_2$H$^+$ distribution.

\subsection{H$_2$CO Excitation Temperatures\label{temp}}

HD~163296 is very favorable for H$_2$CO and N$_2$H$^+$ imaging since 
its relatively high luminosity and massive disk puts the CO snow line at 
a large angular distance compared to other disks, enabling us to   
resolve the H$_2$CO and N$_2$H$^+$ emission. 
Without such spatial information, however, we can still obtain a constraint 
on where H$_2$CO emission originates in disks based on the average H$_2$CO 
excitation temperatures. As a gross approximation, H$_2$CO that coexists 
with CO ice should be cold, i.e. present at an excitation temperature 
comparable to the CO freeze-out temperature. To test the viability of
using excitation temperatures to constrain the H$_2$CO distribution we
extracted spectra from three of the HD~163296 simulations presented in
\S3.2, selecting models with the best outer radius, no inner hole and
power-law indices of $-2$, $0$, and $2$. These distributions
approximately correspond to a H$_2$CO abundance that follows the H$_2$
column, that keeps constant with radius and that increases steeply,
forming a ring. Assuming LTE,  that both H$_2$CO transitions trace the
same underlying populations, and a single rotational excitation
temperature, $T_{\rm rot}$, we calculate  
\begin{equation}
T_{\rm rot}=\frac{E_{\rm1}-E_{\rm 0}}{ln\left((\nu_1 S\mu_1^2  \int T_{\rm 0}dv)/(\nu_0 S\mu_0^2  \int T_{\rm 1}dv)\right)},
\end{equation}
where $E_{\rm0}$ and $E_{\rm 1}$ are the upper energy levels for the 
low and high H$_2$CO transitions used in the calculation 
(H$_2$CO $3_{0,3}-2_{0,2}$ and $4_{1,4}-3_{1,3}$ for most disks), 
$\nu$ and $S\mu^2$ the corresponding line frequencies and temperature 
independent transition strengths and dipole moments,
$\int T dv$ the integrated line intensity, which is calculated from the 
integrated fluxes based on $F/T=13.6\lambda^2/(a\times b)$, where $F$ is 
the flux in Jy, $T$ the intensity in K, $\lambda$ the line wavelength in
millimeters, and $a$ and $b$ the emission diameters in $\arcsec$.
Because of both vertical and radial temperature gradients in the disk,
the size dependence of the emission regions from two transitions is 
complicated but the emission area is not expected to be very
different. For simplicity we assume the extent of the emission is the
same for both transitions on account of the model dependent effects of
the temperature gradients. 
All of the line parameters were gathered from Splatalogue 
(a transition-resolved compilation of several spectroscopic databases), 
with the data originating from CDMS \citep{Muller05}. LTE is a
reasonable approximation if H$_2$CO is mainly present at high
densities.  The critical densities for the observed H$_2$CO
transitions vary between 10$^5$ and 5$\times$10$^6$ cm$^{-3}$
\citep{Troscompt09}, dependent on assumed kinetic temperatures. At
radii $<$300~AU, all gas colder than 25 K is at densities higher than
10$^7$~cm$^{-3}$ \citep{Qi11}, justifying this assumption. 

Using the simulated spectra we derive an excitation temperature of
27~K for the model with $p=-2$ and excitation temperatures close to or
below 20~K for the other two models. Excitation temperatures thus
provide some constraints on the H$_2$CO distribution, but a
temperature close to the expected CO freeze-out temperature only
implies an increasing abundance with radius; it cannot be used to
assess how steeply the abundance increases. 

\begin{figure}[htp]
\epsscale{1.0}
\plotone{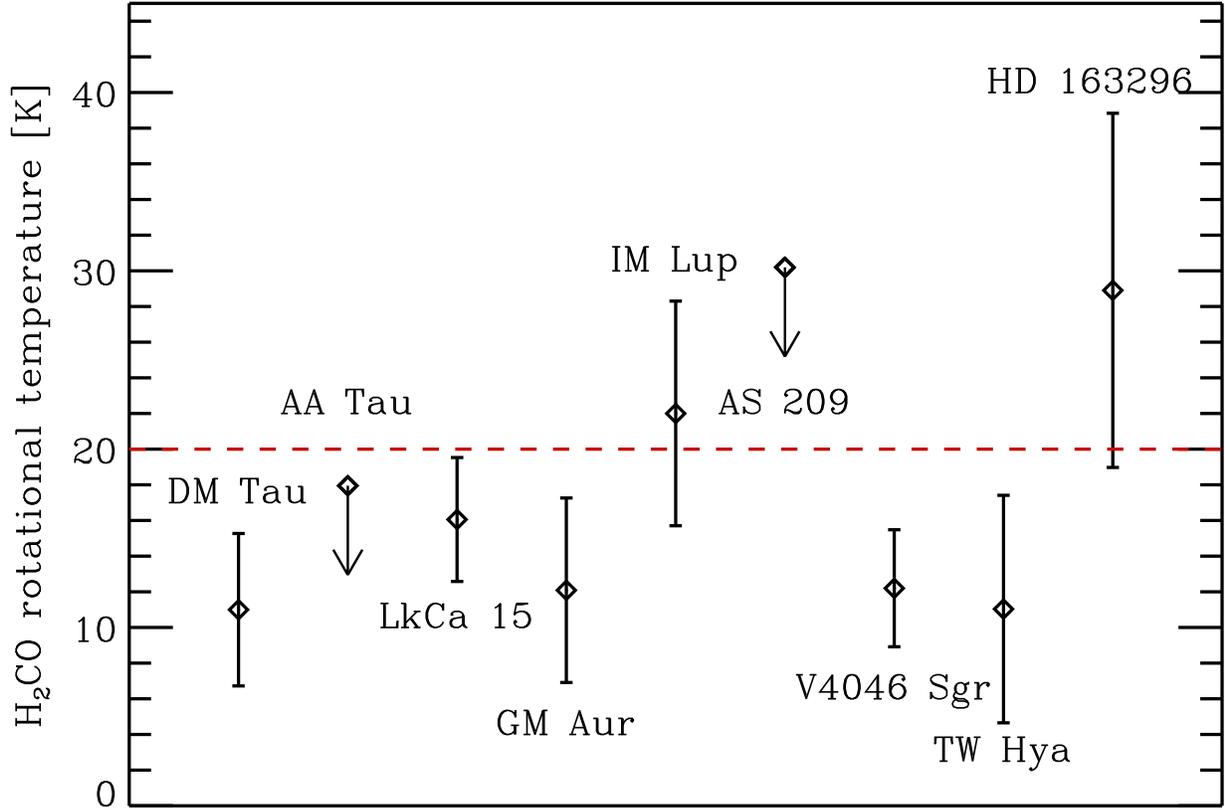}
\caption{The calculated H$_2$CO excitation temperatures for all disks observed with the SMA that have at least one H$_2$CO detection and one upper limit, except for the disk around HD 142527, which has a very high excitation temperature of $>$100~K. All remaining disks have excitation temperatures   H$_2$CO consistent with or lower than the expected CO freeze-out temperature of $\sim$20~K (red dashed horizontal line). \label{fig:temp}}
\end{figure}

By combining the new H$_2$CO detections toward TW~Hya and HD~163296 with 
H$_2$CO data from DISCS \citep{Oberg10c,Oberg11a}, we have a sample of 
10 disks with two H$_2$CO line detections or one H$_2$CO line detection
and one upper limit (Table \ref{tab:star}). These data are sufficient to 
calculate excitation temperatures, albeit with substantial uncertainties. 
Figure \ref{fig:temp} shows the calculated excitation temperatures for 
9 of the 10 disks; the chemically peculiar Herbig Ae star HD 142527 is not 
included because its excitation temperature of 250~K is probably not
due to thermal excitation.  

The H$_2$CO excitation temperature, listed in Table \ref{tab:star},
is consistent with, or lower than, the CO freeze-out temperature of 
$\sim$20~K for all of these disks.
The average H$_2$CO temperature in the sample (excluding HD~142527) 
is $18\pm6$~K. 
We note that the relatively high excitation temperature toward HD~163296 
is most likely due to the fact that some H$_2$CO 3--2 emission is
resolved out by the SMA observations, as the compact-north antenna  
configuration has few short baselines.  

\subsection{H$_2$CO/N$_2$H$^+$ Correlations\label{corr}}

We examine the disk sources to test if N$_2$H$^+$ and H$_2$CO emission 
are correlated across the sample, as would be expected if 1) the two molecules form under 
the same physio-chemical conditions, i.e. only in the regions where CO has 
frozen out, 2) the line emission trace the total N$_2$H$^+$ and
H$_2$CO  column well, and 3) midplane ionization levels and CO
hydrogenation efficiencies do not differ `too much' across the sample
(i.e. the size of the CO freeze-out region is the most important
regulator of N$_2$H$^+$ and H$_2$CO column across the sample). It
should be noted that there are other scenarios that could produce a
correlation as well, and that the correlation analysis below should
only be considered as a constraint on the H$_2$CO distribution in
combination with the results in the previous sections. In particular a
constant H$_2$CO/N$_2$H$^+$ across a disk sample would be expected if
the relative fractions of chemically characteristic disk regions is
always similar in disks. To conclusively test this requires a larger
sample than currently available, but the fact that we did not find
that H$_2$CO emission correlates with any other molecular emission
than N$_2$H$^+$ already challenges this scenario. 

\begin{figure}[htp]
\epsscale{1.0}
\plotone{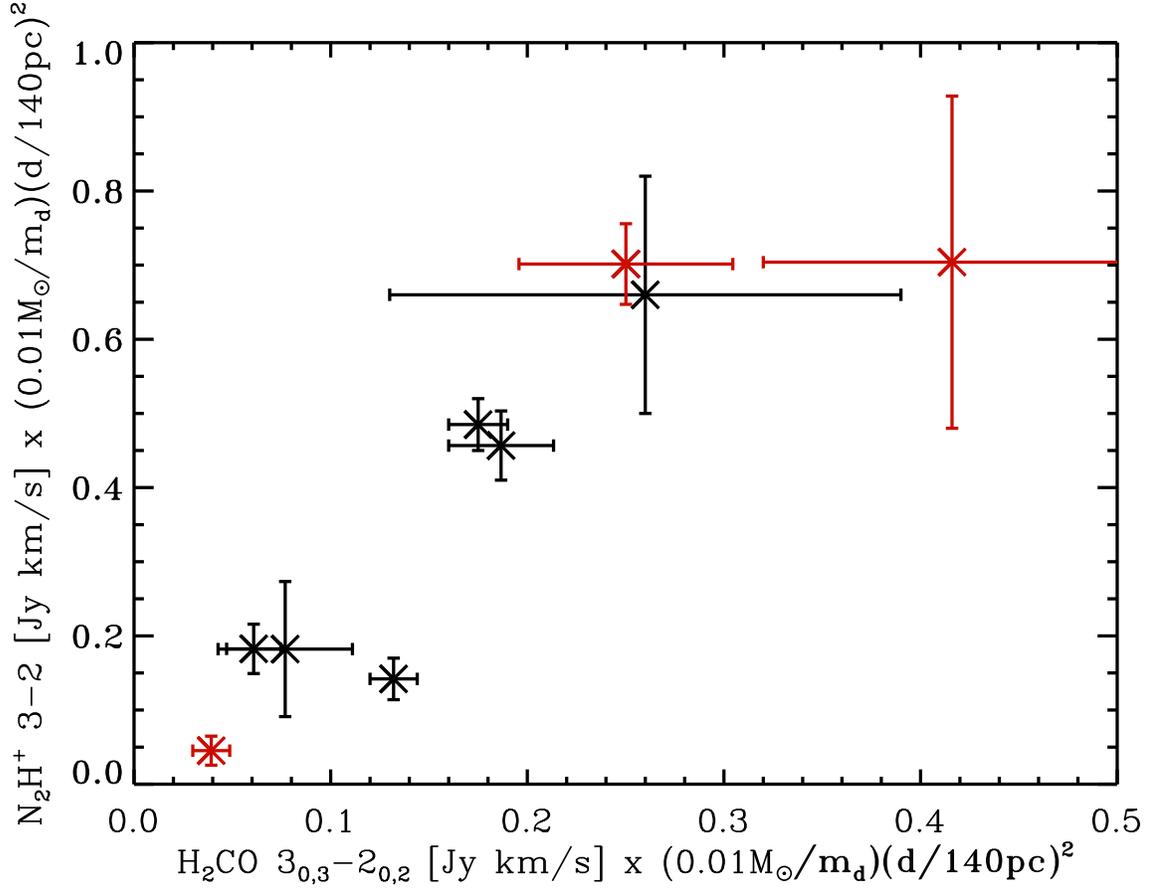}
\caption{Correlation of N$_2$H$^+$ 3--2 (E$_{\rm u}=27$ K) and H$_2$CO 3$_{0,3}-2_{0,2}$ (E$_{\rm u}=21$ K) emission  normalized to disk mass (based on dust modeling) and source distance. The red symbols mark disks where the H$_2$CO 3$_{0,3}-2_{0,2}$  flux has been calculated based on the flux from other H$_2$CO transitions.  \label{fig:corr}}
\end{figure}

\begin{figure}[htp]
\epsscale{1.0}
\plotone{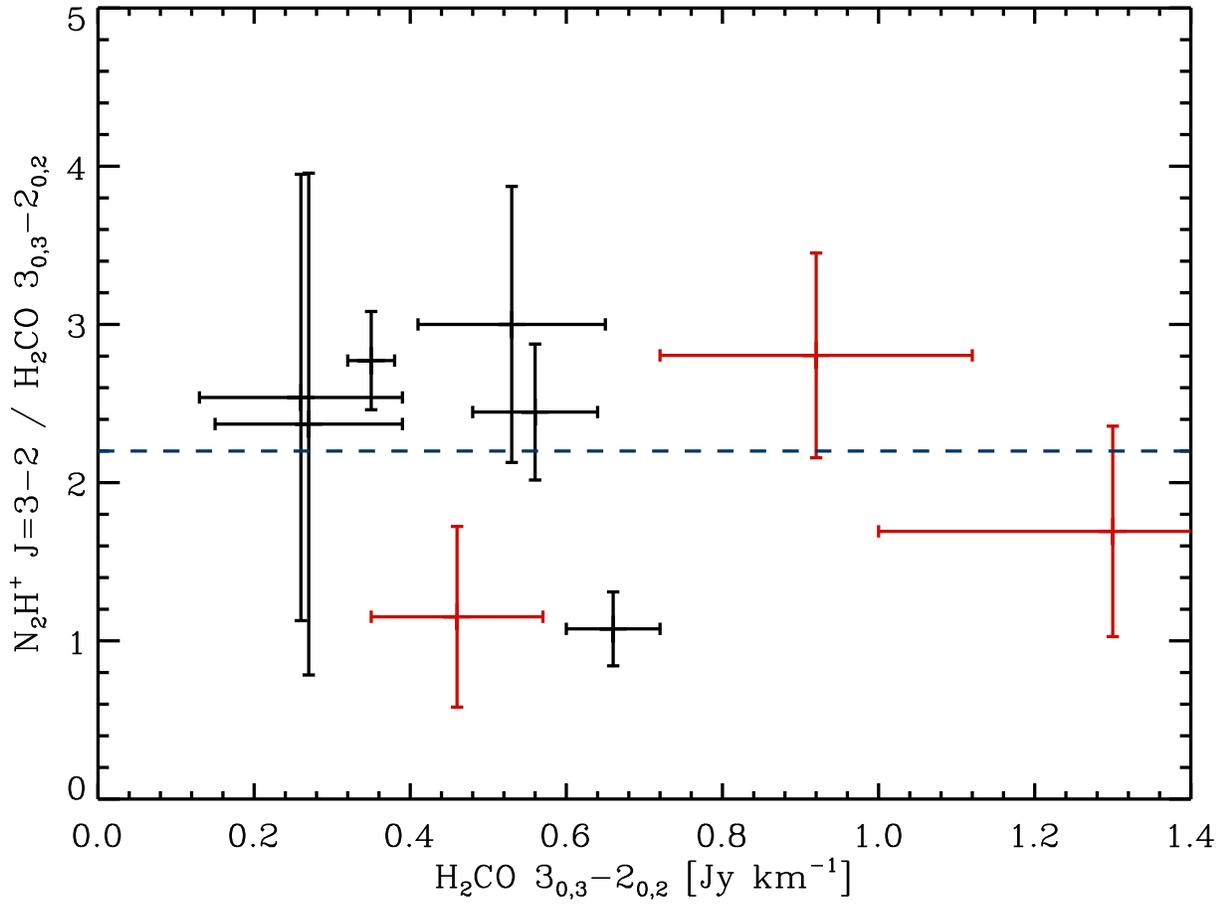}
\caption{ The N$_2$H$^+$/H$_2$CO ratio, demonstrating that it is almost constant, as would be expected from the strong correlation between the normalized fluxes.  \label{fig:corr2}}
\end{figure}

Where possible, we base the comparison on the H$_2$CO $3_{0,3}-2_{0,2}$ line 
that has $E_{\rm up}=21$~K, similar to N$_2$H$^+$ $J=3-2$ ($E_{\rm up}=27$~K),
to minimize variations in fluxes caused by the different detailed temperature 
structures in different disks. Excluding HD 142527, this line has been
observed toward 6/9 of the 
sample. For the remaining 3/9 disks, we calculate the expected 
H$_2$CO $3_{0,3}-2_{0,2}$ line flux based on the H$_2$CO excitation 
temperatures and fluxes of other H$_2$CO lines toward each source. 
We then normalize the flux of each H$_2$CO $3_{0,3}-2_{0,2}$ line 
to the (Taurus) distance of 140~pc, and we further normalize to a disk mass 
of 0.01~M$_\odot$ to account for the fact that more nearby and more massive 
disks tend to have overall stronger line emission. 
Figure \ref{fig:corr} shows that there is a strong correlation between 
the normalized H$_2$CO and N$_2$H$^+$ fluxes in the disk sample; 
the rank correlation is statistically significant at the 95\% level. 
As expected, Figure \ref{fig:corr2} shows that this implies a nearly constant 
N$_2$H$^+$ 3-2 / H$_2$CO 3$_{0,3}-2_{0,2}$ flux ratio across the sample.

\section{Discussion\label{sec:disc}}

Our modeling strategy in this study and in \citet{Qi08} has been to first 
constrain the overall structure of molecular emission in disks using a 
parametric model with a minimum of free parameters, i.e. to determine 
whether the radial column density profile of a species decreases, increases 
or is flat as a function of disk radius. 
As demonstrated in \citet{Oberg12a}, the slope of the radial column density 
profile already can place significant constraints on the formation pathway 
of a molecule. Here we find that H$_2$CO toward HD~163296 belongs to the 
family of molecules that display an increasing column density with radius.
This first-order constraint on the H$_2$CO distribution motivated us to 
consider H$_2$CO formation pathways that would result in an increase of 
H$_2$CO with disk radius. H$_2$CO formation through CO-ice hydrogenation 
results in a simple prediction that H$_2$CO should be present in a ring, 
with the inner edge at the CO snow line. We therefore set up a second model,
based on this prediction, to test if the observations are consistent with 
this hypothesis for H$_2$CO formation. We propose that this combination of 
backward and forward modeling both provides a fair view of the constraints 
obtained by fitting the data, and challenges our basic understanding of disk
chemistry.  

\subsection{H$_2$CO Formation}

H$_2$CO can form through multiple chemical pathways. We have shown that the H$_2$CO distribution towards HD~163296 and the sample statistics are consistent with formation through 
{\it in situ} CO ice hydrogenation. Here we consider the effects of 
additional pathways for H$_2$CO formation, in particular (1) {\it in situ} 
gas phase formation, and (2) formation in the pre- and proto-stellar phases, 
followed by incorporation into the disk. 

H$_2$CO can form in the gas-phase through ion-neutral reactions involving, 
e.g. CH$_3^+$ or through neutral-neutral reactions between CH$_3$ and O 
\citep[e.g.][]{Aikawa99}. The neutral-neutral formation pathway is expected 
to result in a radially flat column density structure for a typical 
T Tauri disk \citep{Aikawa99}, with most emission originating at temperatures 
of 20-40~K \citep{Aikawa03}. It is not clear from existing disk models
whether the structure will look substantially different if the
neutral-ion reactions dominate. In a protostellar chemistry model
\citep{Bergin97}, CH$_3^+$ disappears when CO depletes (E. Bergin,
private communication). This suggests that H$_2$CO forming through
this pathway should be anti-correlated with CO-freeze-out. Neither of
these gas-phase pathways thus 
predicts excess H$_2$CO column densities in the outer disk, or at low 
(T$<$20~K) temperatures. The observed low excitation temperature could
on its own be explained by efficient turbulent mixing of H$_2$CO
formed in the gas-phase and then cooled down in the midplane regions,
similarly to what has been proposed to explain cold CO gas in disks
\citep{Aikawa07}. Turbulent mixing would not, however, explain the
observed deficiency of H$_2$CO towards the inner disk in HD~163296.
Gas-phase formation of H$_2$CO then seems 
an unlikely dominant source of H$_2$CO in disks in light of the new
observations, but the case is unlikely to be conclusively settled
until the exclusive grain-surface product CH$_3$OH is observed to
display a similar distribution. 

H$_2$CO in disks could be a product of protostellar or molecular cloud 
chemistry, as H$_2$CO is commonly observed in pre- and protostellar sources, 
and this molecular content may be preserved, at least in part, through the
process of disk formation. \citet{Willacy07} has modeled this scenario, 
starting with a H$_2$CO ice abundance of 10$^{-6}n_{\rm H}$ inherited 
from the cold cloud. The model also includes H$_2$CO formation through
gas-phase processes in the disk and results are presented with and without
photodesorption. In the model without photodesorption, H$_2$CO follows
CO in the  
inner disk and has an additional outer disk component with an abundance that 
decreases with radius beyond the CO snow line. When photodesorption is
included, the H$_2$CO column is flat across the disk, corresponding to
a power-law index of 0. Neither predicted abundance  
pattern is consistent with the new observations. In addition, H$_2$CO is 
common towards protostellar sources of a range of luminosities 
\citep[e.g.][]{Schoier04, Bisschop07}, while it is pre-dominantly detected 
towards disks around the low luminosity T Tauri stars. 

In short, H$_2$CO formation through {\it in situ} CO-ice hydrogenation 
is not only consistent with the observations, but it is the {\it only} 
pathway proposed (so far) that naturally explains the observations. To
conclusively demonstrate a CO-ice hydrogenation origin would, however,
require the detection of co-spatial emission of CH$_3$OH; CH$_3$OH has
no known efficient gas-phase formation pathway and is predicted to
form together with H$_2$CO whenever CO ice is hydrogenated
\citep{Cuppen09}. 

\subsection{Locating the CO ``Snow Line''}

The ``snow line'' is typically used to denote the midplane disk radius
at which  
the temperature is low enough for water to condense out on dust grains. 
Outside of the snow line, grain accretion will be faster because of larger 
and stickier grains, which may substantially speed up the formation of 
planetesimals and eventually planets \citep[e.g.][]{Hayashi81,Ida04,Ida08,Ciesla06,Kretke07}.
In the Solar System, the dividing line between rocky planets and gas
giants coincide with the H$_2$O snow line \citep{Lewis74}. CO is another abundant volatile in disks and its snow line may boost planet 
formation in the outer disk by providing extra solid masses 
\citep{DodsonRobinson09} and inducing planet traps \citep{Masset06,Hasegawa12}, and could affect the elemental make-up of the 
forming gas-giants \citep{Oberg11e}. Because of its high volatility 
($T_{\rm freeze-out}\sim20$~K), the CO snow line is expected at disk radii 
of 10s--100s of AU. This should make it a far more accessible target than the H$_2$O snow line for millimeter interferometry studies aimed at examining the general effects of snow lines on disk structures. 

\begin{figure}[htp]
\vspace{-1.1in}
\epsscale{1.0}
\plotone{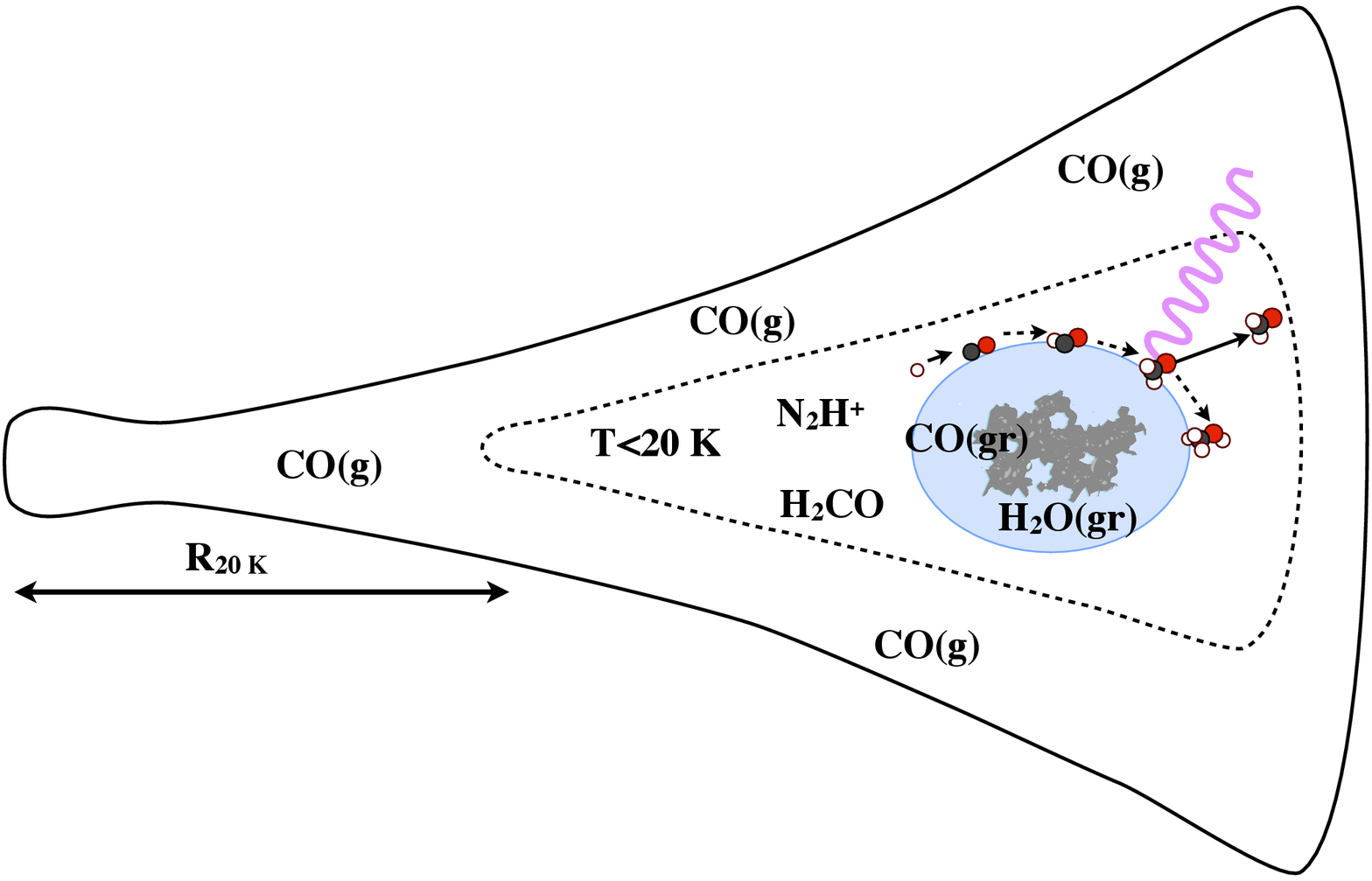}
\vspace{-1.25in}
\caption{Illustration of the expected distribution of CO in the gas-phase and on grain-surfaces. Co freeze-out beyond the CO-snow line ($\rm R_{20K}$) and interior to the dashed contour is predicted to result in a large increase of N$_2$H$^+$ and the onset of H$_2$CO production from CO ice. The ice may then non-thermally desorb to produce gas-phase H$_2$CO. \label{fig:cartoon}}
\end{figure}

Localizing the CO snow line directly from millimeter CO data is 
challenging, however. Disks have both a radial temperature gradient
away from the central star, and a vertical one set by radiative
heating at the disk surface \citep[e.g.][]{Aikawa99}. This results 
in a CO condensation front that is located at different radii at 
different disk heights (Fig. \ref{fig:cartoon}), and also that some 
CO is present at all radii in the upper disk layers. This fact, 
together with a complex radiative transfer (most CO lines are expected 
to be optically thick in the disk center and optically thin in the outer
parts of the disk), means that the location of a CO snow line in a disk 
cannot be inferred from simply inspecting a CO disk image.  The best
constraint that exists to date on a CO snow line radius is toward the Herbig Ae
star HD~163296 based on the analysis of multi-transition, multi-isotope, 
spatially resolved CO line data on a self-consistent physical disk model
\citep{Qi11}. The temperature structure of the model has been
constrained by optically thick multiple CO lines and detailed
analysis of the optically thinner $^{13}$CO emission reveals a
significant column density reduction at around 165 AU that cannot be
explained by the overall disk column distribution as traced by the
dust. This is interpreted as the result of CO freeze-out and the
location as the CO snow line. Uncertainties in the temperature
structure will mainly affect the determination of the CO freeze-out
temperature at the location of the CO snow line, rather than the
location itself , which is 
constrained between 135 and 175~AU in the disk of HD~163296. 
While fruitful, this is a time consuming approach that will be difficult
to apply to larger samples of disks and will always include some
degree of model dependency.

Another approach to constrain the CO snow line location is to identify
trace species that are only present where CO  has begun to freeze
out. Such molecules should display a ring-like structure with the
inner edge corresponding to the midplane CO snow line. 
As we have described, N$_2$H$^+$ and H$_2$CO are good candidate probes 
of CO snow lines. In principle, these species can be used as powerful 
chemical imaging tools to constrain CO snow line locations in large 
samples of disks, rather than relying on complex analysis of the CO 
isotopologue observations.  

\subsubsection{Simulated ALMA Observations \label{sec:alma}}

It seems clear that both N$_2$H$^+$ and H$_2$CO are outer-disk species
from the chemical perspective. To connect their formation to the onset of 
CO freeze-out conclusively requires a combination of higher sensitivity, 
and higher resolution imaging. This kind of imaging can be 
done with the newly available capabilities of the ALMA telescope in 
Chile, which is nearing completion of construction.

\begin{figure*}[htp]
\epsscale{1.0}
\plotone{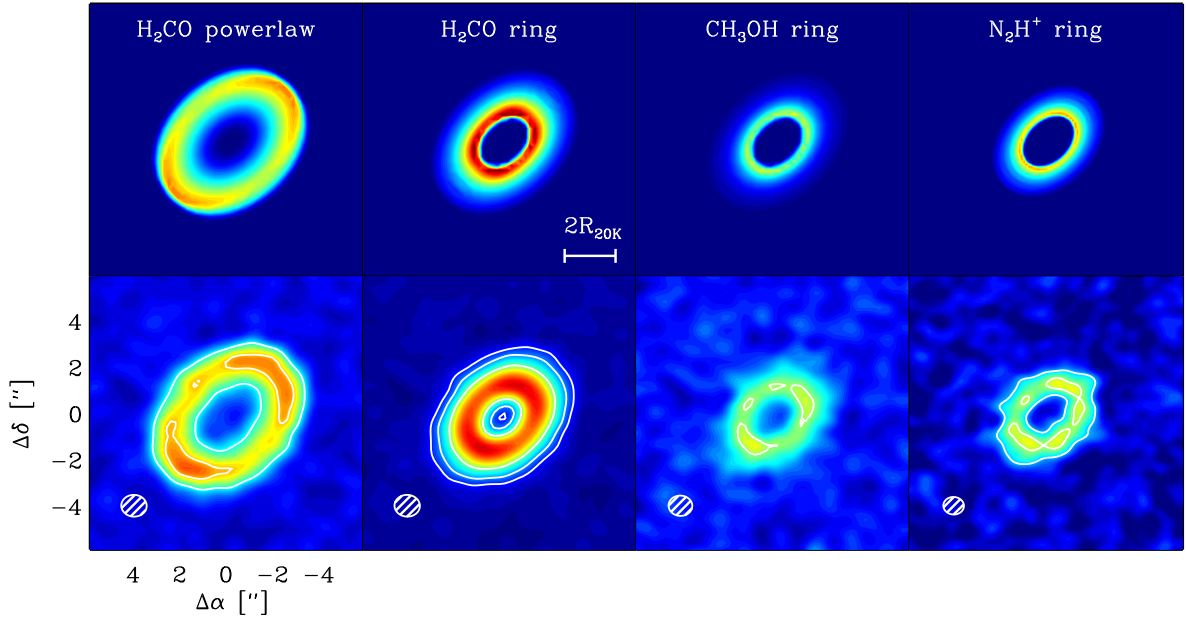}
\caption{The predicted morphology of H$_2$CO 3--2 emission toward
  HD~163296 for the power-law and ring model, when using ALMA antenna
  configuration 5 (corresponding to 0.3\arcsec~resolution) and 1h integration. The H$_2$CO ring structure should be accompanied by similar CH$_3$OH and N$_2$H$^+$ rings if the observations reported in this paper are due the exclusive presence of H$_2$CO in the outer disk, exterior to the CO snow line.  For the CH$_3$OH simulation we assumed the same column density and excitation temperature for CH$_3$OH as has been observed for H$_2$CO, and imaged the 241.767 GHz 5$_{0,5}-4_{0,4}$ line. White contours are [0.025,0.05,0.1] Jy km s$^{-1}$ per beam. \label{fig:sim}}
\end{figure*}

To demonstrate the astrochemical predictions generated by our analysis of 
the SMA data, and the ease at which they can be tested with ALMA, we present 
a set of simulations of HD~163296 ALMA observations using the antenna
configuration 5 (corresponding to 0.3\arcsec~resolution) 
in Figure \ref{fig:sim}. The predicted H$_2$CO and N$_2$H$^+$ rings are
readily detected at this resolution, and the power-law and ring models 
are clearly distinguished. For both models, the diameter of the emission 
maximum can be estimated directly from the images within a fraction of the 
beam size, enabling us to determine if the H$_2$CO and N$_2$H$^+$ emission 
truly trace the CO snow line. In addition, ALMA should have sufficient 
sensitivity to detect CH$_3$OH if it is present with a similar abundance 
and distribution as H$_2$CO, as expected if both of these species are formed 
through CO hydrogenation and then non-thermally desorbed. 

\section{Conclusions}

We have presented three observational results that support the idea that
CO freeze-out regulates the H$_2$CO and N$_2$H$^+$ chemistry in disks:

\begin{enumerate}
\item H$_2$CO and N$_2$H$^+$ emission towards HD~163296 appears offset from 
the continuum peak at a size scale consistent with the CO snow line at 160~AU. 
These observations are matched well by a ring model where H$_2$CO is present 
only in the disk regions with CO freeze-out.
\item The H$_2$CO excitation temperature in a sample of 9 disks is typically 
below $\sim20$~K, consistent with the bulk of H$_2$CO emission originating 
in disk regions where CO is expected to freeze-out. 
\item The H$_2$CO and N$_2$H$^+$ emission are correlated across the disk 
sample, consistent with the hypothesis of coexistence beyond the CO snow line.
\end{enumerate}

These results suggest that both N$_2$H$^+$ and H$_2$CO should be present in rings, with the inner edge at the CO snow line. This may be used as a probe of CO snow line locations across samples of disk and is also important for predicting the organic content of comets forming at different disk radii. In general, the radial and vertical distributions of molecules constitute 
strong probes of the basic chemistry used into astrochemical models, while 
molecular abundances and column densities are probably best used to test our 
understanding of the structure and history of individual objects. 

{\it Facilities:} \facility{SMA}

\acknowledgments

\noindent  The SMA is a joint project between the Smithsonian
Astrophysical Observatory and the Academia Sinica Institute of
Astronomy and Astrophysics and is funded by the Smithsonian
Institution and the Academia Sinica. We thank Edwin Bergin and
Paola~D'Alessio for their helpful suggestions, and a referee for
constructive comments on the paper.
Support for K.~I.~O. is provided by NASA through a Hubble Fellowship
grant  awarded by the Space Telescope Science Institute, which is
operated by the Association of Universities for Research in Astronomy,
Inc., for NASA, under contract NAS 5-26555. We also acknowledge NASA
Origins of Solar Systems grant No. NNX11AK63.  

\bibliographystyle{aa}

\begin{thebibliography}{56}
\expandafter\ifx\csname natexlab\endcsname\relax\def\natexlab#1{#1}\fi

\bibitem[{{Aikawa} \& {Herbst}(1999)}]{Aikawa99}
{Aikawa}, Y. \& {Herbst}, E. 1999, \aap, 351, 233

\bibitem[{{Aikawa} {et~al.}(2003){Aikawa}, {Momose}, {Thi}, {van Zadelhoff},
  {Qi}, {Blake}, \& {van Dishoeck}}]{Aikawa03}
{Aikawa}, Y., {Momose}, M., {Thi}, W., {et~al.} 2003, \pasj, 55, 11

\bibitem[{{Aikawa} \& {Nomura}(2006)}]{Aikawa06}
{Aikawa}, Y. \& {Nomura}, H. 2006, \apj, 642, 1152

\bibitem[{{Aikawa}(2007)}]{Aikawa07}
{Aikawa}, Y. 2007, \apjl, 656, L93

\bibitem[{{Aikawa} {et~al.}(2002){Aikawa}, {van Zadelhoff}, {van Dishoeck}, \&
  {Herbst}}]{Aikawa02}
{Aikawa}, Y., {van Zadelhoff}, G.~J., {van Dishoeck}, E.~F., \& {Herbst}, E.
  2002, \aap, 386, 622

\bibitem[{{Andrews} {et~al.}(2011){Andrews}, {Wilner}, {Espaillat}, {Hughes},
  {Dullemond}, {McClure}, {Qi}, \& {Brown}}]{Andrews11}
{Andrews}, S., {Wilner}, D., {Espaillat}, C., {et~al.} 2011, ArXiv e-prints

\bibitem[{{Andrews} \& {Williams}(2005)}]{Andrews05}
{Andrews}, S.~M. \& {Williams}, J.~P. 2005, \apj, 631, 1134

\bibitem[{{Andrews} {et~al.}(2009){Andrews}, {Wilner}, {Hughes}, {Qi}, \&
  {Dullemond}}]{Andrews09}
{Andrews}, S.~M., {Wilner}, D.~J., {Hughes}, A.~M., {Qi}, C., \& {Dullemond},
  C.~P. 2009, \apj, 700, 1502

\bibitem[{{Bergin} \& {Langer}(1997)}]{Bergin97}
{Bergin}, E.~A. \& {Langer}, W.~D. 1997, \apj, 486, 316

\bibitem[{{Bergin} {et~al.}(2002){Bergin}, {Alves}, {Huard}, \&
  {Lada}}]{Bergin02}
{Bergin}, E.~A., {Alves}, J., {Huard}, T., \& {Lada}, C.~J. 2002, \apjl, 570,
  L101

\bibitem[{{Bisschop} {et~al.}(2007){Bisschop}, {J{\o}rgensen}, {van Dishoeck},
  \& {de Wachter}}]{Bisschop07}
{Bisschop}, S.~E., {J{\o}rgensen}, J.~K., {van Dishoeck}, E.~F., \& {de
  Wachter}, E.~B.~M. 2007, \aap, 465, 913

\bibitem[{{Caselli} {et~al.}(1999){Caselli}, {Walmsley}, {Tafalla}, {Dore}, \&
  {Myers}}]{Caselli99}
{Caselli}, P., {Walmsley}, C.~M., {Tafalla}, M., {Dore}, L., \& {Myers}, P.~C.
  1999, \apjl, 523, L165

\bibitem[{{Ciesla} \& {Cuzzi}(2006)}]{Ciesla06}
{Ciesla}, F.~J. \& {Cuzzi}, J.~N. 2006, \icarus, 181, 178

\bibitem[{{Cuppen} {et~al.}(2009){Cuppen}, {van Dishoeck}, {Herbst}, \&
  {Tielens}}]{Cuppen09}
{Cuppen}, H.~M., {van Dishoeck}, E.~F., {Herbst}, E., \& {Tielens}, A.~G.~G.~M.
  2009, \aap, 508, 275

\bibitem[{{Dodson-Robinson} {et~al.}(2009){Dodson-Robinson}, {Willacy},
  {Bodenheimer}, {Turner}, \& {Beichman}}]{DodsonRobinson09}
{Dodson-Robinson}, S.~E., {Willacy}, K., {Bodenheimer}, P., {Turner}, N.~J., \&
  {Beichman}, C.~A. 2009, \icarus, 200, 672

\bibitem[{{Dutrey} {et~al.}(1997){Dutrey}, {Guilloteau}, \&
  {Guelin}}]{Dutrey97}
{Dutrey}, A., {Guilloteau}, S., \& {Guelin}, M. 1997, \aap, 317, L55

\bibitem[{{Dutrey} {et~al.}(2007){Dutrey}, {Henning}, {Guilloteau}, {Semenov},
  {Pi{\'e}tu}, {Schreyer}, {Bacmann}, {Launhardt}, {Pety}, \&
  {Gueth}}]{Dutrey07}
{Dutrey}, A., {Henning}, T., {Guilloteau}, S., {et~al.} 2007, \aap, 464, 615

\bibitem[{{Fuchs} {et~al.}(2009){Fuchs}, {Cuppen}, {Ioppolo}, {Romanzin},
  {Bisschop}, {Andersson}, {van Dishoeck}, \& {Linnartz}}]{Fuchs09}
{Fuchs}, G.~W., {Cuppen}, H.~M., {Ioppolo}, S., {et~al.} 2009, \aap, 505, 629

\bibitem[{{Garrod} {et~al.}(2007){Garrod}, {Wakelam}, \& {Herbst}}]{Garrod07}
{Garrod}, R.~T., {Wakelam}, V., \& {Herbst}, E. 2007, \aap, 467, 1103

\bibitem[{{Hasegawa} \& {Pudritz}(2012)}]{Hasegawa12}
{Hasegawa}, Y. \& {Pudritz}, R.~E. 2012, ArXiv e-prints

\bibitem[{{Hayashi}(1981)}]{Hayashi81}
{Hayashi}, C. 1981, in IAU Symposium, Vol.~93, Fundamental Problems in the
  Theory of Stellar Evolution, ed. D.~{Sugimoto}, D.~Q. {Lamb}, \& D.~N.
  {Schramm}, 113--126

\bibitem[{{Henning} \& {Semenov}(2008)}]{Henning08}
{Henning}, T. \& {Semenov}, D. 2008, in IAU Symposium, Vol. 251, IAU Symposium,
  ed. S.~{Kwok} \& S.~{Sandford}, 89--98

\bibitem[{{Hogerheijde} \& {van der Tak}(2000)}]{Hogerheijde00}
{Hogerheijde}, M.~R. \& {van der Tak}, F.~F.~S. 2000, \aap, 362, 697

\bibitem[{{Ida} \& {Lin}(2004)}]{Ida04}
{Ida}, S. \& {Lin}, D.~N.~C. 2004, \apj, 616, 567

\bibitem[{{Ida} \& {Lin}(2008)}]{Ida08}
{Ida}, S. \& {Lin}, D.~N.~C. 2008, \apj, 685, 584

\bibitem[J{\o}rgensen(2004)]{Jorgensen04} J{\o}rgensen, J.~K.\ 2004, \aap, 424, 589 

\bibitem[{{Kretke} \& {Lin}(2007)}]{Kretke07}
{Kretke}, K.~A. \& {Lin}, D.~N.~C. 2007, \apjl, 664, L55

\bibitem[{{Lewis}(1974)}]{Lewis74}
{Lewis}, J.~S. 1974, Science, 186, 440

\bibitem[{{Lombardi} {et~al.}(2008){Lombardi}, {Lada}, \& {Alves}}]{Lombardi08}
{Lombardi}, M., {Lada}, C.~J., \& {Alves}, J. 2008, \aap, 489, 143

\bibitem[{{Masset} {et~al.}(2006){Masset}, {Morbidelli}, {Crida}, \&
  {Ferreira}}]{Masset06}
{Masset}, F.~S., {Morbidelli}, A., {Crida}, A., \& {Ferreira}, J. 2006, \apj,
  642, 478

\bibitem[{{M{\"u}ller} {et~al.}(2005){M{\"u}ller}, {Schl{\"o}der}, {Stutzki},
  \& {Winnewisser}}]{Muller05}
{M{\"u}ller}, H.~S.~P., {Schl{\"o}der}, F., {Stutzki}, J., \& {Winnewisser}, G.
  2005, Journal of Molecular Structure, 742, 215
  
\bibitem[{{Mumma} \& {Charnley}(2011)}]{Mumma11}
{Mumma}, M.~J. \& {Charnley}, S.~B. 2011, \araa, 49, 471

\bibitem[{{{\"O}berg} {et~al.}(2009{\natexlab{a}}){{\"O}berg}, {Linnartz},
  {Visser}, \& {van Dishoeck}}]{Oberg09c}
{{\"O}berg}, K.~I., {Linnartz}, H., {Visser}, R., \& {van Dishoeck}, E.~F.
  2009{\natexlab{a}}, \apj, 693, 1209

\bibitem[{{{\"O}berg} {et~al.}(2011{\natexlab{a}}){{\"O}berg}, {Murray-Clay},
  \& {Bergin}}]{Oberg11e}
{{\"O}berg}, K.~I., {Murray-Clay}, R., \& {Bergin}, E.~A. 2011{\natexlab{a}},
  \apjl, 743, L16

\bibitem[{{{\"O}berg} {et~al.}(2010){{\"O}berg}, {Qi}, {Fogel}, {Bergin},
  {Andrews}, {Espaillat}, {van Kempen}, {Wilner}, \& {Pascucci}}]{Oberg10c}
{{\"O}berg}, K.~I., {Qi}, C., {Fogel}, J.~K.~J., {et~al.} 2010, \apj, 720, 480

\bibitem[{{{\"O}berg} {et~al.}(2011{\natexlab{b}}){{\"O}berg}, {Qi}, {Fogel},
  {Bergin}, {Andrews}, {Espaillat}, {Wilner}, {Pascucci}, \&
  {Kastner}}]{Oberg11a}
{{\"O}berg}, K.~I., {Qi}, C., {Fogel}, J.~K.~J., {et~al.} 2011{\natexlab{b}},
  \apj, 734, 98

\bibitem[{{{\"O}berg} {et~al.}(2011{\natexlab{c}}){{\"O}berg}, {Qi}, {Wilner},
  \& {Andrews}}]{Oberg11d}
{{\"O}berg}, K.~I., {Qi}, C., {Wilner}, D.~J., \& {Andrews}, S.~M.
  2011{\natexlab{c}}, \apj, 743, 152

\bibitem[{{{\"O}berg} {et~al.}(2012){{\"O}berg}, {Qi}, {Wilner}, \&
  {Hogerheijde}}]{Oberg12a}
{{\"O}berg}, K.~I., {Qi}, C., {Wilner}, D.~J., \& {Hogerheijde}, M.~R. 2012,
  ApJ, 749, 162

\bibitem[{{{\"O}berg} {et~al.}(2005){{\"O}berg}, {van Broekhuizen}, {Fraser},
  {Bisschop}, {van Dishoeck}, \& {Schlemmer}}]{Oberg05}
{{\"O}berg}, K.~I., {van Broekhuizen}, F., {Fraser}, H.~J., {et~al.} 2005,
  \apjl, 621, L33

\bibitem[{{{\"O}berg} {et~al.}(2009{\natexlab{b}}){{\"O}berg}, {van Dishoeck},
  \& {Linnartz}}]{Oberg09b}
{{\"O}berg}, K.~I., {van Dishoeck}, E.~F., \& {Linnartz}, H.
  2009{\natexlab{b}}, \aap, 496, 281

\bibitem[{{Pinte} {et~al.}(2008){Pinte}, {Padgett}, {M{\'e}nard},
  {Stapelfeldt}, {Schneider}, {Olofsson}, {Pani{\'c}}, {Augereau},
  {Duch{\^e}ne}, {Krist}, {Pontoppidan}, {Perrin}, {Grady}, {Kessler-Silacci},
  {van Dishoeck}, {Lommen}, {Silverstone}, {Hines}, {Wolf}, {Blake}, {Henning},
  \& {Stecklum}}]{Pinte08}
{Pinte}, C., {Padgett}, D.~L., {M{\'e}nard}, F., {et~al.} 2008, \aap, 489, 633

\bibitem[{{Qi} {et~al.}(2011){Qi}, {D'Alessio}, {{\"O}berg}, {Wilner},
  {Hughes}, {Andrews}, \& {Ayala}}]{Qi11}
{Qi}, C., {D'Alessio}, P., {{\"O}berg}, K.~I., {et~al.} 2011, \apj, 740, 84

\bibitem[{{Qi} {et~al.}(2004){Qi}, {Ho}, {Wilner}, {Takakuwa}, {Hirano},
  {Ohashi}, {Bourke}, {Zhang}, {Blake}, {Hogerheijde}, {Saito}, {Choi}, \&
  {Yang}}]{Qi04}
{Qi}, C., {Ho}, P.~T.~P., {Wilner}, D.~J., {et~al.} 2004, \apjl, 616, L11

\bibitem[Qi et al.(2003)]{Qi03} Qi, C., Kessler, J.~E., 
Koerner, D.~W., Sargent, A.~I., \& Blake, G.~A.\ 2003, \apj, 597, 986 

\bibitem[{{Qi} {et~al.}(2008){Qi}, {Wilner}, {Aikawa}, {Blake}, \&
  {Hogerheijde}}]{Qi08}
{Qi}, C., {Wilner}, D.~J., {Aikawa}, Y., {Blake}, G.~A., \& {Hogerheijde},
  M.~R. 2008, \apj, 681, 1396

\bibitem[{{Qi} {et~al.}(2006){Qi}, {Wilner}, {Calvet}, {Bourke}, {Blake},
  {Hogerheijde}, {Ho}, \& {Bergin}}]{Qi06}
{Qi}, C., {Wilner}, D.~J., {Calvet}, N., {et~al.} 2006, \apjl, 636, L157

\bibitem[Rodriguez et al.(2010)]{Rodriguez10} Rodriguez, D.~R., 
Kastner, J.~H., Wilner, D., \& Qi, C.\ 2010, \apj, 720, 1684 


\bibitem[{{Sch{\"o}ier} {et~al.}(2004){Sch{\"o}ier}, {J{\o}rgensen}, {van
  Dishoeck}, \& {Blake}}]{Schoier04}
{Sch{\"o}ier}, F.~L., {J{\o}rgensen}, J.~K., {van Dishoeck}, E.~F., \& {Blake},
  G.~A. 2004, \aap, 418, 185

\bibitem[{{Sch{\"o}ier} {et~al.}(2005){Sch{\"o}ier}, {van der Tak}, {van
  Dishoeck}, \& {Black}}]{Schoier05}
{Sch{\"o}ier}, F.~L., {van der Tak}, F.~F.~S., {van Dishoeck}, E.~F., \&
  {Black}, J.~H. 2005, \aap, 432, 369

\bibitem[{{Thi} {et~al.}(2004){Thi}, {van Zadelhoff}, \& {van
  Dishoeck}}]{Thi04}
{Thi}, W., {van Zadelhoff}, G., \& {van Dishoeck}, E.~F. 2004, \aap, 425, 955

\bibitem[{{Tielens} \& {Hagen}(1982)}]{Tielens82}
{Tielens}, A.~G.~G.~M. \& {Hagen}, W. 1982, \aap, 114, 245

\bibitem[{{Troscompt}{et~al.}(2009){Troscompt},{Faure},{Wiesenfeld},{Ceccarelli}, \& {Valiron}}]{Troscompt09}
{Troscompt}, N. and {Faure}, A. and {Wiesenfeld}, L. and {Ceccarelli}, C. and 
	{Valiron}, P. 2009, \aap, 493, 687

\bibitem[{{Verhoeff} {et~al.}(2011){Verhoeff}, {Min}, {Pantin}, {Waters},
  {Tielens}, {Honda}, {Fujiwara}, {Bouwman}, {van Boekel}, {Dougherty}, {de
  Koter}, {Dominik}, \& {Mulders}}]{Verhoeff11}
{Verhoeff}, A.~P., {Min}, M., {Pantin}, E., {et~al.} 2011, \aap, 528, A91

\bibitem[{{Walsh} {et~al.}(2012){Walsh}, {Nomura}, {Millar}, \&
  {Aikawa}}]{Walsh12}
{Walsh}, C., {Nomura}, H., {Millar}, T.~J., \& {Aikawa}, Y. 2012, \apj, 747,
  114

\bibitem[{{Watanabe} {et~al.}(2003){Watanabe}, {Shiraki}, \&
  {Kouchi}}]{Watanabe03}
{Watanabe}, N., {Shiraki}, T., \& {Kouchi}, A. 2003, \apjl, 588, L121

\bibitem[{{Willacy}(2007)}]{Willacy07}
{Willacy}, K. 2007, \apj, 660, 441

\end{thebibliography}

\newpage

\begin{table*}
\caption{Observational Parameters$^a$}
{\scriptsize
\begin{tabular}{lcccc}
\hline\hline
 Parameter & \multicolumn{3}{c}{HD~163296} \\
\cline{2-4} 
 & 2007 Mar 20 (COM-N) & 2012 Jun 10 (COM) & 2012 Jun 10 (COM)\\
 &                  &                 & 2012 Aug 12 (SUB) \\
 &                  &                 & 2012 Aug 14 (SUB) \\
\smallskip
Baseline (m) & 16.4--139.2 & 16.4--77.0 & 9.5--77.0\\
\hline
Lines & H$_2$CO 3$_{1,2}$--2$_{1,1}$  & H$_2$CO 4$_{1,4}$--3$_{1,3}$ & N$_2$H$^+$ 3--2       \\
\hline
Rest frequency (GHz) & 225.69778 & 281.52693 & 279.51170 \\
Beam Size (FWHM) & 3\farcs4$\times$2\farcs2 & 3\farcs9$\times$2\farcs5 
                 & 4\farcs9$\times$3\farcs4 \\
P.A. & 54$^\circ$ & 15$^\circ$ & 14$^\circ$ \\
Channel spacing (km\,s$^{-1}$) & 0.54 & 0.87 & 0.60 \\
RMS Noise (Jy\,beam$^{-1}$) & 0.063 & 0.096 & 0.072\\
\smallskip
Integrated Flux (Jy\,km\,s$^{-1}$) & 0.89 & 1.55 & 0.53 \\
\hline
Continuum (GHz) & 221 & 273 & 273 \\
\hline
Beam Size (FWHM) & 3\farcs5$\times$2\farcs2 & 4\farcs1$\times$2\farcs6
                 & 5\farcs2$\times$3\farcs5 \\
P.A. & 56$^\circ$ & 15$^\circ$ & 14$^\circ$ \\
RMS Noise (mJy\,beam$^{-1}$) & 3.4 & 9.4 & 3.8 \\
Flux Density (Jy\,beam$^{-1}$) & 0.49 & 0.96 & 0.95 \\
Integrated Flux (Jy) & 0.61 & 1.06 & 1.07 \\
\hline\hline
 Parameter & \multicolumn{3}{c}{TW~Hya} \\
\cline{2-4} 
&2012 Jan 13(SUB)& 2008 Feb 23 (COM)& 2012 Jun 4 (COM) \\
\smallskip
Baseline (m) & 9.5--45.2 &16.4--77.0 & 16.4--77.0 \\
\hline
Lines &  H$_2$CO 4$_{1,4}$--3$_{1,3}$     & H$_2$CO 5$_{1,5}$--4$_{1,4}$ & N$_2$H$^+$ 3--2      \\
\hline
Rest frequency (GHz) & 281.52693    & 351.76866 &  279.51170 \\ 
Beam Size (FWHM) & 9\farcs1$\times$5\farcs6 
                 & 3\farcs3$\times$2\farcs0 & 3\farcs5$\times$2\farcs0 \\
P.A. & 24$^\circ$ & 1$^\circ$ &
    10$^\circ$ \\
Channel spacing (km\,s$^{-1}$) & 0.43 & 0.17 &
    0.11 \\
RMS Noise (Jy\,beam$^{-1}$) &  0.16 & 0.23 & 0.55 \\
\smallskip
Integrated Flux (Jy\,km\,s$^{-1}$) & 1.22 & 0.54 & 2.2 \\
\hline
Continuum (GHz) &  273 & 346 & 273 \\
\hline
Beam Size (FWHM) & 9\farcs2$\times$5\farcs7
                 & 3\farcs5$\times$2\farcs0 & 4\farcs8$\times$2\farcs4\\
P.A. &  25$^\circ$ & -2$^\circ$ & 10$^\circ$ \\
RMS Noise (mJy\,beam$^{-1}$) &  6.2 & 12 & 14 \\
Flux Density (Jy\,beam$^{-1}$) &  0.90 & 1.20 & 0.70 \\
Integrated Flux (Jy) & 0.93 & 1.52 & 0.92 \\
\hline
\end{tabular}
}
\tablenotetext{a}{All quoted values assume natural weighting.}
\label{tab:obs}
\end{table*}

\begin{table}
\caption{Best-Fit Model Parameters$^{\rm a}$}
\begin{center}
\begin{tabular}{l c}
\hline
Parameter & HD~163296 \\
\hline
\hline
\multicolumn{2}{c}{Power-law Model} \\
\hline
$\rm N_{100}$ (cm$^{-2}$) & 1.7$\times$10$^{11}$ \\
$\rm p$ & 2 \\
$\rm R_{out}$ (AU) & 410 \\
$\rm R_{in}$ (AU) & $<$200 \\
{\it Surface boundary $\sigma_s$} & {\it 10$^{-0.1}$=0.79} \\
Midplane boundary $\sigma_m$ & 10$^{1.5}$=31.6 \\
$\chi^2$ & 720101 \\
\hline
\multicolumn{2}{c}{Ring Model} \\
\hline
Fractional abundance & 5.5$\times10^{-11}$ \\
{\it $R_{out}$ (AU)} & {\it 500} \\
{\it Surface boundary} & {\it T$<$19 K} \\
Midplane boundary $\sigma_m$ & 10$^{1.5}$=31.6 \\
$\chi^2$ & 720101 \\
\hline
\end{tabular}
\end{center}
\label{tab:best-fit}
$^{\rm a}$Parameters in italics were fixed from CO modeling
($\sigma_s$ for the power-law model and $R_{out}$ for the ring model).
\end{table}

\begin{deluxetable}{lcccccccc}
\tabletypesize{\scriptsize}
\tablecaption{Central star and disk, and H$_2$CO data. \label{tab:star}}
\tablewidth{0pt}
\tablehead{
\colhead{Source} & \colhead{RA} & \colhead{DEC} & \colhead{Spec.} & \colhead{d} & \colhead{m$_{disk}$} &\colhead{Detected H$_2$CO$^{\rm i}$}&\colhead{Calc. $3_{0,3}-2_{0,2}$} &\colhead{T$_{\rm rot}$}\\
&  &  & \colhead{type}  & \colhead{[pc]} & \colhead{ [M$_{\odot}$]} &\colhead{}&\colhead{[Jy km s$^{-1}$]} &\colhead{[K]}
}
\startdata

TW~Hya$^{\rm a}$    	& 11 01 51.91 	& $-$34 42 17.0 	&K7 	&56		&0.005 & $4_{1,4}-3_{1,3}$, $5_{1,5}-4_{1,4}$ &$\sim$1.3$^{\rm j}$ &11(6)\\
\smallskip
HD~163296$^{\rm b}$& 17 56 21.29	& $-$21 57 21.9 	&A1	&122	&0.089 &$3_{1,2}-2_{1,1}$, $4_{1,4}-3_{1,3}$ &$\sim$0.46$^{\rm j}$ &29(10)\\
DM~Tau$^{\rm c}$     	& 04 33 48.73  	& $+$18 10 10.0 	&M1	&140	&0.02 &$3_{0,3}-2_{0,2}$, $4_{1,4}-3_{1,3}$ & -- &11(4)\\
AA Tau$^{\rm c}$ 	& 04:34:55.42  	& $+$24:28:53.2 	&K7	&140	&0.01&$3_{0,3}-2_{0,2}$, ($4_{1,4}-3_{1,3}$)$^{\rm k}$ & -- &$<$18\\
LkCa 15$^{\rm c}$   	& 04 39 17.78  	& $+$22 21 03.5 	&K5	&140	&0.05 &$3_{0,3}-2_{0,2}$, $4_{1,4}-3_{1,3}$ & -- & 16(3)\\
GM Aur$^{\rm c}$  	& 04 55 10.98  	& $+$30 21 59.4 	&K3	&140	&0.03  &$3_{0,3}-2_{0,2}$, $4_{1,4}-3_{1,3}$ & -- & 12(5)\\
IM Lup $^{\rm d,e}$ 	&15 56 09.23  	&  $-$37 56 05.9 	&M0 &150	&0.08 &$3_{0,3}-2_{0,2}$, $4_{1,4}-3_{1,3}$ & -- & 22(6)\\
AS 209$^{\rm f}$  	&16 49 15.29  	& $-$14 22 08.6  	&K5	&125	&0.028 &$3_{0,3}-2_{0,2}$, ($4_{1,4}-3_{1,3}$)$^{\rm k}$ & -- & $<$30\\
V4046 Sgr$^{\rm g}$&18 14 10.47  	& $-$32 47 34.5 	&K5	&73		&0.01 &$3_{1,2}-2_{1,1}$, $4_{1,4}-3_{1,3}$ &$\sim$0.92$^{\rm j}$ & 12(3)\\
HD 142527$^{\rm h}$&15 56 41.89  & $-$42 19 23.3	&F6	&145	&0.1 &$3_{0,3}-2_{0,2}$, $4_{1,4}-3_{1,3}$ & -- & 250(120)\\
\enddata
\\Star and disk data from: $^{\rm a}$\citet{Qi04}, $^{\rm b}$\citet{Qi11}, $^{\rm c}$\citet{Andrews05}, $^{\rm d}$\citet{Lombardi08}, $^{\rm e}$\citet{Pinte08}, $^{\rm f}$\citet{Andrews09},  $^{\rm g}$\citet{Rodriguez10},  $^{\rm h}$\citet{Verhoeff11}.\\
$^{\rm i}$From this work and \citet{Oberg10c,Oberg11a}. Upper limits in parentheses. $^{\rm j}$Calculated from other H$_2$CO transition strengths and the determined excitation temperature. $^{\rm k}$The $3-2$ lines were reported as upper limits in \citet{Oberg10c,Oberg11a}, but after re-examination of the emission maps, we conclude that these lines are better described as tentative detections (2-3$\sigma$).
\end{deluxetable}

\end{document}